\documentclass{ws-ijmpa}

\usepackage[T1]{fontenc}
\usepackage[latin1]{inputenc}
\usepackage[english]{babel}
\usepackage{amsmath}
\usepackage{latexsym}
\usepackage{graphicx}

\usepackage{amsfonts}
\usepackage{amssymb}
\usepackage{dsfont}

\usepackage{epsfig}

\def\beq{\begin{equation}}
\def\eeq{\end{equation}}
\def\bitem{\begin{itemize}}
\def\eitem{\end{itemize}}
\def\bear{\begin{array}}
\def\ear{\end{array}}

\def\munu{{\mu\nu}}

\def\d{\partial}

\def\dr{\d_{\rho}}

\def\m{\hat{m}}

\begin{document}

\markboth{Lacquaniti Montani}
{On Matter coupling in 5D KK model}

%
\catchline{}{}{}{}{}
%

\title{On matter coupling in 5D Kaluza-Klein model}
\author{Valentino Lacquaniti$^{1,2,3}$, Giovanni Montani$^{1,4,5}$}

\address{1- ICRA---International Center for Relativistic Astrophysics, 
Physics Department (G9),
University  of Rome, "`La Sapienza"', 
Piazzale Aldo Moro 5\\ Rome, I-00185, Italy.\\ valentino.lacquaniti@icra.it, montani@icra.it \\
 2- Physics Department  "`E.Amaldi "`,  University of Rome, "`Roma Tre"', 
Via della Vasca Navale 84\\  Rome, I-00146, Italy \\ 
 3-LAPTH -9, Chemin de Bellevue BP 110 74941\\ Annecy Le Vieux Cedex, France \\
 4-ENEA- C. R. Frascati ( Department F. P. N. ), via E.Fermi 45\\ I-00044, Frascati, Rome, Italy \\
 5-ICRANet - C.C. Pescara, Piazzale della Repubblica 10\\  I-65100, Pescara, Italy }

\maketitle

\begin{history}
\received{Day Month Year}
\revised{Day Month Year}
\end{history}

\begin{abstract}
We analyze some unphysical features of the geodesic approach to matter coupling in a compactified Kaluza-Klein scenario, like the q/m puzzle and the huge massive modes. We  propose a new approach, based on Papapetrou multipole expansion, that provides a new equation for the motion of a test particle. We show how this equation provides right couplings and does not generate huge massive modes. 
\keywords{Kaluza-Klein; Matter Coupling }
\end{abstract}

\ccode{PACS numbers: }

\section*{}	
The standard approach to the  test particle dynamics  in General Relativity relies in the equivalence between the geodesic trajectory, driven by the Action $S_{g}=\int \,ds$, and the physical motion of the particle, governed by the Action
$S=-mS_g$.
Thus, the simplest choice to study a 5D test particle in Kaluza-Klein ( KK ) model (\cite{kk},\cite{phi})  is to generalize this approach;  therefore we assume that the dynamics is driven by the Action
\beq
S_5=-\tilde{m}\int ds_5 \quad \quad  ds_5^2=ds^2-\phi^2(ekA_{\mu}dx^{\mu}+dx^5).
\label{azgeod5}
\eeq
Here $ds_5$ is the 5D line element, $A_{\mu}$ is the electromagnetic field, $\phi$ the extra scalar field allowed in modern  KK theories (\cite{phi},\cite{lm}), and $ek$ is a dimensional constant that can be related to the present observed value of $G$ via $4G_{obs}=(ek)^2\phi^2c^4$. Finally, $\tilde{m}$ is an unknown constant mass parameter: an important  point, indeed, is to determine whether $\tilde{m}$ represents the physical rest mass of the reduced 4D particle.
The first step is, however, to calculate the Eulero-Lagrange equations associated to the Action (\ref{azgeod5}), and $\tilde{m}$ will not appear in the outcome.  After calculation we have a set of five equations:
\beq
\frac{dw_5}{ds}=0 ,\quad\quad
\frac{Du^{\mu}}{Ds}=F^{\munu}u_{\nu}\left( \frac {ekw_5}{\sqrt{1+\frac{w_5^2}{\phi^2}}}\right)+(u^{\mu}u^{\nu}-g^{\munu})\frac{\d_{\nu}\phi}{\phi^3}\left(      \frac {w_5}{\sqrt{1+\frac{w_5^2}{\phi^2}}}           \right)^2.
\label{oldeq}
\eeq
where $F^{\munu} $ is the Faraday tensor, $\frac{D}{Ds}$ is the covariant derivative along the path and $w_5=J_{A5}\frac{dx^A}{ds_5}$ ( where $J_{AB}$ is the 5D metrics and $A,B=0,1,2,3,5$), $u^{\mu}=\frac{dx^{\mu}}{ds}$. Hence we have $w_5$ as a constant of motion from the first equation, and moreover it  is possible to show  that $w_5$ is a scalar; thus, other equation describes the motion of a 4D particle and we can recognize the electrodynamics coupling, in terms of the charge and the physical rest mass, once we define $\frac{q}{mc^2}= ekw_5(1+\frac{w_5^2}{\phi^2})^{-\frac12}$. Unfortunately, recalling the value of $ek$, we get the following bound:
$$
\frac{q^2}{4Gm^2}=\frac{w_5^2}{\phi^2}\frac{1}{1+\frac{w_5^2}{\phi^2}}<1.
$$
Such a bound is violated by every known elementary particle. Therefore, although the geodesic approach provides an equation with the correct structure, it yields unphysical coupling. 
The problem of huge massive modes is strictly linked to the $q/m$ puzzle. Indeed, by considering the Action (\ref{azgeod5}) and addressing the Hamiltonian procedure, we get the conjugate momenta and the 5D dispersion relation $P_AP^A=\tilde{m}^2$. Via the canonical quantization, we get a 5D Klein-Gordon equation whose Lagrangian reads $\mathcal{L}=J^{AB}(\d_{A}\zeta)(\d_B\zeta)^+-\tilde{m}^2\zeta\zeta^+$, where a generic complex scalar field  ( as toy model ) is concerned. Now, considering a KK field, i.e. $\zeta(x^{\mu},x^5)=\eta(x^{\mu})e^{ip_5x^5}$, where $p_5$ is scalar and conserved, the reduced Lagrangian reads:
\beq
\mathcal{L}=g^{\munu}(-i\d_{\mu}-ekp_5A_{\mu})\eta[(-i\d_{\nu}-ekp_5A_{\nu})\eta)]^+-(\tilde{m}^2+\frac{p_5^2}{\phi^2}).
\eeq
We recognize a U(1) gauge invariant Lagrangian where the reduced field $\eta$ acquires  a charge $ekp_5$ and a mass term $m^2=(\m^2+\frac{p_5^2}{\phi^2})$; the ratio $q/m$  fits to the result previously obtained for the motion of the test particle, but is clear now that  $\m$ does not represent the correct rest mass for the particle. Hence $\m$ is an unknown 5D mass whose physical meaning is not clearly established within this scheme. Moreover, requiring the compactness of the fifth dimension, we get  the quantization of $p_5$, and so on of the charge. Thus, the discretization of $p_5$ gives rise to a tower of modes for the mass term $m$;  fixing the minimum value of $p_5$ via the elementary charge $e$ we get the extra dimension size below our observational limit, but, at the same way, we get huge massive modes beyond the Planck scale. Therefore, the problem of the charge-mass ratio is strictly linked to the problem of massive modes. Indeed, the puzzle of matter coupling is an historical problem of KK models (\cite{matterproblem}).
\\
We have to admit that the 5D geodesic approach does not provide the correct dynamics for the particle, i.e. the trajectory of the 5D particle is no more the 5D geodesic. Our idea is to face the problem of the test particle and its rest mass, within the multipole expansion scheme of Papapetrou (\cite{pap}), that is more general than the geodesic approach. Then, given a generic 5D matter tensor $T^{AB}$, we state the following model:
$$
D_AT^{AB}=0,\quad\d_5T^{AB}=0.
$$
The first equation is the conservation law, while  the second equation is given for consistency with the cylindricity hypothesis. The KK reduction of the first equation yields the following set:
\beq
 \nabla_{\mu} j^{\mu}=0, \quad\quad
 \nabla_{\rho}T^{\mu\rho}=(\frac{\dr\phi}{\phi})T^{\mu\rho}+g^{\mu\rho}(\frac{\dr\phi}{\phi^3})T_{55}+\phi^3F^{\mu\rho} j_{\rho} .
\eeq
where
$ j_{\mu}=-ek\frac{T_5^{\mu}}{\phi^3}$.
Now,  we assume that  matter tensor is peaked in a thin tube centered around a 4D trajectory $X^{\mu}$ and negligible  outside. This is consistent with the cylindricity hypothesis, i.e. the unobservability of extra-dimension, and with the phenomenological request that we observe trajectories only in our 4D space. Then, after we integrate over the space, we perform a Taylor expansion of center $X^{\mu}$ , i.e. $x^{\mu}=X^{\mu}+\delta x^{\mu}$. At the lowest order the equation of motion reads
 \beq
\frac{D}{Ds}(mu^{\mu})=\frac23\frac{d}{ds}\left(\frac{1}{\phi^3}\right)\phi^3mu^{\mu}+g^{\mu\rho}(\frac{\dr\phi}{\phi^3})A+qF^{\mu\rho}u_{\rho},
\label{moto1}
\eeq
where $ds^2=g_{\munu}dX^{\mu}dX^{\nu}$, $u^{\mu}=\frac{dX^{\mu}}{ds}$,  $\frac{D}{Ds}$ is the covariant derivative along the path, and we have defined the scalar quantities $m$, $q$, $A$ :
\beq
m=\frac{1}{u^0}\int\!\!\!d^3x\,\sqrt{g}\,\frac{T^{00}}{\phi^3} \quad\quad q=-ek\int\!\!\!d^3x\,\sqrt{g}\,\frac{T_5^0}{\phi^3}\quad \quad A=u^0\int\!\!\!d^3x\,\sqrt{g}\, \frac{T_{55}}{\phi^3}
\eeq
By  requiring  $u_\mu\frac{Du^{\mu}}{Ds}=0$ we find the new relevant condition
\beq
\frac{dm}{ds}=\left(\frac{A}{\phi^3}-\frac{2m}{\phi}\right)\frac{d\phi}{ds}.
\label{massvar}
\eeq
and finally we can recast the equation of motion as follows, with  no explicit derivative of mass: 
\beq
m\frac{Du^{\mu}}{Ds}=A(g^{\mu\rho}-u^{\rho}u^{\mu})\frac{\d_{\rho}\phi}{\phi^3}+qF^{\mu\rho}u_{\rho}.
\label{neweq}
\eeq
Let us now compare the old equation (\ref{oldeq}) to this new one (\ref{neweq}). We recognize that i) they show the same dynamical structure but coupling factors are not the same ii) in (\ref{neweq}) we have three factors that are defined in terms of independent degrees of freedom of the matter tensor , therefore are not correlated each to other, while in the geodesic approach $q$ and $m$ were both defined in terms of $p_5$ ( so giving the upper bound ): therefore now no bound arises iii) q is conserved due to the presence of a conserved current ( gauge theory ); A is not constant but in principle there is no symmetry requiring its conservation iv) mass is not conserved and this is indeed the relevant feature of this new equation. Anyway there is no reason to require in principle the conservation of $m$. Indeed this approach shows that the physical motion in a 5D space is not the geodesic one; the reason is that in 5D KK model the 5D Equivalence Principle is violated (\cite{lm}); therefore mass is not necessarily a constant. The conservation of mass, however can be restored if we take $\phi=1$ or $A=2m\phi^2$. Finally, in support of these observations we shows how this approach does not generate huge massive modes.
The  equation (\ref{neweq}) admits the following Action:
$$
S=-\int\!\!\! m\,d\,s+q(A_{\mu}dx^{\mu}+dx^{5}),
$$
where $m$ has to be regarded as a variable function whose derivative  reads 
$$ 
\d_{\mu}m=\frac{\d_{\mu}\phi}{\phi^3}A+\frac23 u_{\mu}\frac{d}{ds}\left(\frac{1}{\phi^3}\right)\phi^3m.
$$
After calculating Lagrangian, Hamiltonian and momenta we have:
$$
P_AP^A=m^2-\frac{q^2}{\phi^2} \quad \Rightarrow \quad  g^{\munu}\Pi_{\mu}\Pi_{\nu}=m^2,
$$
where
$ P_5=q$, $ \Pi_{\mu}=mu_{\mu}=P_{\mu}-qA_{\mu}$. Therefore,  in the resulting Klein-Gordon equation  we now  have the  counter term $-\frac{q^2}{\phi^2}$ that rules out the huge massive modes. In our opinion this approach, even if  not definitive, provides a simple solution for the historical matter problem in KK model, without removing the hypothesis of the compactification. Thus, from a theoretical point of view it enforces the physical meaning of KK theories and deserves a detailed investigation, also in view of its multidimensional extension. Promising perspectives appear the search of correct currents associated with gauge symmetries in multidimensional KK models, the analysis of the complete cosmological solution with matter, and the comparison to models for dark energy
with mass varying particles (\cite{amendola}).

\end{document}